\documentclass[amsmath,english,amssymb,12pt]{iopart}

\usepackage{graphicx}
\usepackage{psfrag}

\begin{document}
\title{Rate of Homogeneous Crystal Nucleation in molten NaCl}

\author{C. Valeriani$^1$, E. Sanz$^2$ and D. Frenkel$^1$}

\address{1 FOM Institute for Atomic
  and Molecular Physics,\\
  Kruislaan 407, 1098 SJ Amsterdam, The Netherlands\\ }
\address{2 Facultad de Ciencias Qu\'imicas,\\
Universidad Complutense de Madrid,\\
28040 Madrid, Spain\\}
\date{\today}
\begin{abstract}

We report a numerical simulation of the rate of crystal nucleation
of sodium chloride from its melt at moderate supercooling. In this
regime nucleation is too slow to be studied with ``brute-force''
Molecular Dynamics simulations. The melting temperature of
(``Tosi-Fumi'') NaCl is $\sim 1060$K. We studied crystal
nucleation at $T$=$800$K and $825$K. We observe that the critical
nucleus formed during the nucleation process has the crystal
structure of bulk NaCl. Interestingly, the critical nucleus is
clearly faceted: the nuclei have a cubical shape. We have computed
the crystal-nucleation rate using two completely different
approaches, one based on an estimate of the rate of diffusive
crossing of the nucleation barrier, the other based on the Forward
Flux Sampling and Transition Interface Sampling (FFS-TIS) methods.
We find that the two methods yield the same result to within an
order of magnitude. However, when we compare the extrapolated
simulation data with the only available experimental results for
NaCl nucleation, we observe a discrepancy of nearly 5 orders of
magnitude.  We discuss the possible causes for this discrepancy.

\end{abstract}


\section{Introduction}
Crystallization of salts is a phenomenon of great practical
relevance. In fact, it is one of the most important industrial
separation processes. But it also plays a crucial role in
geological processes that occur on an altogether different time
scale. The crystallization process consists of two steps:
nucleation and growth. If nucleation is slow compared to the time
it takes a crystal to grow to a size comparable to the size of the
container, large single crystals will form (an example is rock
salt). When nucleation is fast, the resulting solid will form as a
fine powder. It is clearly important to be able to predict the
rate of nucleation of salts and -- at a later stage -- to
understand the factors that influence nucleation. In the present
paper we aim to demonstrate that, with current simulation
techniques and currently available force-fields, it is indeed
possible to compute the rate of nucleation of a real salt crystal
(in the present case NaCl from its melt) This opens the way to
``ab-initio'' predictions of nucleation rates of many ionic
substances.

Solutions or melts can often be cooled well below their freezing
temperature. The reason is that the formation of small nuclei of
the stable crystal phase is an activated process that may be
extremely slow. Intuitively, it is easy to understand why crystal
nucleation is an activated process, i.e. why there is a
free-energy barrier separating the metastable parent phase (the
liquid) from the stable crystal phase. The point is that,
initially, the formation of small crystalline nuclei costs free
energy. But once the crystal nucleus exceeds a critical size, its
free energy decreases as it grows. The rate at which crystal
nuclei form depends strongly on $\Delta G_{crit}$, the free-energy
required to form a critical nucleus. Classical Nucleation Theory
(CNT) is commonly used to estimate the height of the nucleation
barrier and to predict the rate of crystal
nucleation~\cite{Gibbs,Volmer,Farkas,Becker,Kelton}. According to
CNT, the total free energy of a crystallite that forms in a
supersaturated solution or melt contains two terms: the first is a
``bulk'' term that expresses the fact that the solid is more
stable than the supersaturated fluid - this term is negative and
proportional to the volume of the crystallite. The second is a
``surface'' term that takes into account the free-energy cost of
creating a solid-liquid interface. This term is positive and
proportional to the surface area of the crystallite. According to
CNT, the total (Gibbs) free-energy cost to form a spherical
crystallite with radius $R$ is

\begin{equation}
\Delta G=\frac{4\pi}{3} R^{3}\rho _{S}\Delta \mu +4\pi R^{2}\gamma , \label{CNT1}
\end{equation}
where $\rho _{S}$ is the number density of the solid, $\Delta \mu
$ ($<0$) the difference in chemical potential of the solid and the
liquid, and $\gamma$ is the solid-liquid interfacial free energy
density. The function $\Delta G$ goes through a maximum at
$R=2\gamma /(\rho _{S}|\Delta \mu |)$ and the height of the
nucleation barrier is
\begin{equation}
\Delta G_{crit}=\frac{16\pi }{3}\gamma ^{3}/(\rho _{S}|\Delta \mu |)^{2}%
.  \label{CNTBarrier}
\end{equation}
The crystal-nucleation rate per unit volume, $I$, depends strongly
on $\Delta G_{crit}$:
\begin{equation}
I=\kappa \exp (-\Delta G_{crit}/k_{B}T) .
\label{Nucleation Rate}
\end{equation}
Here $\kappa $\ is a kinetic prefactor, $T$\ is the absolute
temperature and $k_{B}$\ is Boltzmann's constant. The CNT
expression for the nucleation rate then becomes
\begin{equation}
I=\kappa\; \exp \left[ -\frac{16\pi}{3}\gamma ^{3}/(\rho_{S}|\Delta \mu |)^{2} \right] .
\label{eqn:rate}
\end{equation}

Under experimental conditions, nucleation is infrequent on the
time scale of typical molecular processes. Yet, when it happens,
it proceeds rapidly.  This makes it difficult to study the
structure and dynamics of crystal nuclei of atoms or small
molecules in experiments. In the case of NaCl, the experiments are
also complicated by the fact that crystallization occurs at high
temperatures.  This may explain why there is a scarcity of
experimental data on the nucleation of NaCl. To our knowledge, the
only data are those of Buckle and
Ubbelhode~\cite{NaClexp11,NaClexp12,NaClexp2} from the 1960's. In these
experiments, crystallization in NaCl micro-droplets was observed
visually. As the droplet size (($\mathcal{O}( 3\mu)$) and time
window for the measurement  (1--30 seconds -- after which the
droplets sedimented out of view) -- were fixed, the nucleation
rate could  be determined at one temperature only (905 K for
NaCl). At this temperature, the nucleation rate was such that, on
average, one nucleus would form during the observation time
($\mathcal{O}(10s)$) in a droplet with a volume of order
$10^{-17}m^3$. Hence the experimental nucleation rate per unit
volume was  $O(10^{16})$ $m^{-3}s^{-1}$.

For experimental nucleation rates of this order of magnitude,
brute-force MD simulations are out of the question. The average
time it would take for nuclei to form spontaneously in a system
consisting of several thousands of particles is of order
of $10^{20}$ seconds. Clearly, this is beyond the scope of MD
simulations. The standard solution to circumvent this problem is
to perform simulations at much larger undercooling than used in
the experiments.  Huang et al. \cite{Bartell} performed MD
simulations of melting and freezing of a droplet composed of 216
NaCl ions in vacuum: to this end,  they performed temperature
quenches down to 550K (i.e. approximately half the melting
temperature)  and found nucleation rates of the order of
$O(10^{36})$$m^{-3}s^{-1}$, which is 20 orders of magnitude higher
than the experimental rate at $905K$.

Another effort to study nucleation at less severe supercooling
was made by  Koishi et al. \cite{Japanese}. These authors
performed an MD simulation of 125000 ions system in vacuum, at
temperatures of  740K (i.e. approximately 0.7 the melting temperature $T_m$)
and 640K (i.e. approximately 0.6 the melting temperature $T_m$). Both free
and periodic boundary conditions were used. The estimated
nucleation rate at $740K$ was $O(10^{35})$$m^{-3}s^{-1}$, which is
virtually the same value as found by Huang et al.~\cite{Bartell}
at a much larger supercooling. This is surprising because
nucleation rates tend to depend very strongly on temperature. This
suggests that, at least at the lowest temperatures, the barrier
for crystal nucleation is negligible. More in general, crystal
nucleation under extreme supercooling need not proceed following
the same path as under moderate supercooling~\cite{Kelton}.

In what follows, we use the technique of
refs.~\cite{VanDuijneveldt,Wolde,Auer} based on a combination of
umbrella sampling~\cite{umbrella} (to determine the barrier
height) and a dynamical simulation (to determine the crossing
rate). The computing time required for this scheme does not scale
exponentially with the nucleation barrier, but it does increase
with increasing nucleus size.

We also compute the nucleation rate using an algorithm based on
the "forward flux sampling" (FFS) ~\cite{FFS} and the "Transition
Path Samplig" (TIS) techniques ~\cite{TIS} and compare it with the
one obtained using the method previously mentioned.

In the present work we study homogeneous crystal nucleation in the
Tosi-Fumi NaCl model for NaCl at two different temperatures, viz.
$T_{1}$=$800$K and $T_{2}$=$825$K, corresponding to $25 \%$ and
$22 \%$ undercooling. For this system, we computed the nucleation
barrier, examined the structure and shape of the critical nucleus
and computed the nucleation rate.

\section{Methods}

The Tosi-Fumi rigid-ion interaction potential for NaCl is of the
following form \cite{Tosi1,Tosi2},
 \begin{equation}
U_{ij}(r)=A_{ij}e^{[B(\sigma_{ij}-r)]}-\frac{C_{ij}}{r^{6}}-\frac{D
        _{ij}}{r^{8}}+\frac{q_i q_j}{r} \ ,
\end{equation}
where the parameters have the values given in table~\ref{tavola3}.
\begin{table}[h!]
  \centerline{
    \begin{tabular}{||c|c|c|c|c|c||}
      \hline {  }          & $A_{ij}$  &  $B$ & $C_{ij}$ & $D_{ij}$     & $\sigma_{ij}$
      \\         & $[kJ/mol]$  &  $[\dot{A} ^{-1}]$ & $[\dot{A}^{6}kJ/mol]$ & $[\dot{A}^{8}kJ/mol]$     & $[\dot{A}]$
      \\ \hline { Na-Na }  &  $25.4435$        &  $3.1546$      & $101.1719$              & $48.1771$    & $2.340$
      \\ \hline { Na-Cl }  &  $20.3548$        &  $3.1546$      & $674.4793$              & $837.0770$   & $2.755$
      \\ \hline { Cl-Cl }  &  $15.2661$        &  $3.1546$      & $6985.6786$             & $14031.5785$ & $3.170$
      \\ \hline
    \end{tabular}}
\caption{Potential parameters for NaCl.}
\label{tavola3}
\end{table}

This pair potential is written as the sum of a Born-Mayer
repulsion, two attractive van der Waals contributions and a
Coulomb interaction term. In our simulations, we calculated the
Coulomb interactions using the Ewald summations method with a real
space cutoff of 10 $\dot{A}$  and a real space damping parameter of
0.25 $\dot{A}^{-1}$.
We truncated the Van Der Waals part of the potential at  9 $\dot{A}$,
assuming the $g(r)=1$ beyond this cutoff.

The computed number density
of ions in the bulk solid at $800$K and $825$K at $10^5$ Pa was
$0.041$$\dot{A}^{-3}$, in agreement with experiment~\cite{Janz}.
The density of ions in the supercooled liquid at the same
temperature and pressure was $0.034$$\dot{A}^{-3}$.

We prepared under cubic boundary conditions a supercooled system of $(12)^3$ NaCl ion pairs at
ambient pressure by cooling it down below the melting temperature.
For the present model, Anwar et al.~\cite{Jamshed} have computed
the melting temperature: $T_{m}$=$(1064 \pm 14)$K, which is very
close to the experimental melting temperature
($T_m^{exp}$=$1072$K). Using constant-pressure Monte Carlo
simulations, we cooled the system down to the temperatures where
we studied nucleation: $T_{1}$=$800$K and $T_{2}$=$825$K,
corresponding to $25 \%$  and $22 \%$ undercooling. Note that the
experiments on NaCl nucleation where performed at a somewhat
higher temperature ($16 \%$ supercooling). The reason why we could
not perform simulations  at these higher temperatures is that the critical
nucleus would be about twice the size of the nucleus that  could
be studied without spurious finite-size effects~\cite{Andersen} for
the system sizes that we employed. At temperatures below 750K,
spontaneous nucleation occurred during the simulations. We
therefore kept the temperature above this lower limit.

Nucleation is an activated  process. In steady-state, the
nucleation rate per unit volume and time is given by eq.
(\ref{Nucleation Rate}) \cite{Volmer,Oxtoby,Kelton}
where $\exp(-\beta \Delta G_{crit})$ is the equilibrium
probability per nucleus to find a critical nucleus in the metastable parent
phase. $\kappa$ is a kinetic prefactor. In the case of a diffusive
barrier crossing, $\kappa$ can be expressed as:
\begin{eqnarray}\label{eqn:prefactor}
\kappa = \sqrt{\frac{|\Delta \mu|}{6 \pi k_{B}T n_{crit}}} \; \rho_{liq}
f^{+}_{n^{crit}}
\end{eqnarray}
where $\rho_{liq}$ is the ion density of the metastable liquid,
$f^{+}_{n^{crit}}$ the rate at which particles are added to a
critical nucleus, $\Delta\mu$=$\mu_{liq}-\mu_{sol}$ is the difference
in chemical potential between liquid and solid and
$\sqrt{\frac{|\Delta \mu|}{6 \pi k_{B}T n_{crit}}}$ the Zeldovitch factor,
that takes into account the fact that, during steady-state
nucleation, the concentration of critical nuclei is, in fact, not
the equilibrium concentration.

It is important to distinguish between $f^{+}_{ncrit}$, which is the 
rate at which particles are added to a  nucleus with the critical size, 
and the net flux across the nucleation barrier. In steady state, this 
net flux is equal the number of nuclei that go from $ncrit$ to $ncrit+1$ 
minus the number that go from $ncrit+1$ to 
$ncrit$~\cite{Volmer,Farkas,Becker}. Hence, the actual nucleation rate 
is a combination of the forward rate $f^{+}$ and the backward rate 
$f^{-}$. However, because of detailed balance, knowledge of 
$f^{+}_{ncrit}$ (combined with knowledge of the barrier height and 
shape) is enough to compute the nucleation rate.

Assuming a  diffusive attachment or detachment of single particles
from the critical nucleus, the forward rate $f^{+}_{crit}$ at the
top of the barrier can be related to the spontaneous fluctuations
in the number of particles in a nucleus at the top of the
nucleation barrier:
\begin{eqnarray}\label{eqn:diff}
f^{+}_{crit} =  \frac{1}{2}   \frac{ \left< \Delta n_{crit}^{2}(t) \right> }{t}
\end{eqnarray}
where $\left< \Delta n_{crit}^{2}(t)\right>$ =
$\left<[n_{crit}(t)-n_{crit}(0)]^{2}\right>$ is the mean square
change in the number of particles belonging to the critical
nucleus during a time interval $t$. To estimate $f^{+}_{crit}$ a
series of dynamical trajectories were necessary: after generating
a set of uncorrelated configurations at the top of the barrier,
we carried out NVT MD simulations using the $DL\_POLY$
package\cite{DLPOLY} with a timestep of 0.5 fs.
We computed the nucleation rate using equation (\ref{Nucleation Rate}).

Moreover, we computed the nucleation rate per unit time and volume
using an algorithm based on path-sampling techniques
~\cite{FFS,TIS} and compared the results with those obtained using
equation (\ref{Nucleation Rate}).

\section{Results}
\begin{figure}[h!]
\begin{center}
  \includegraphics[width=8.5cm,clip=true]{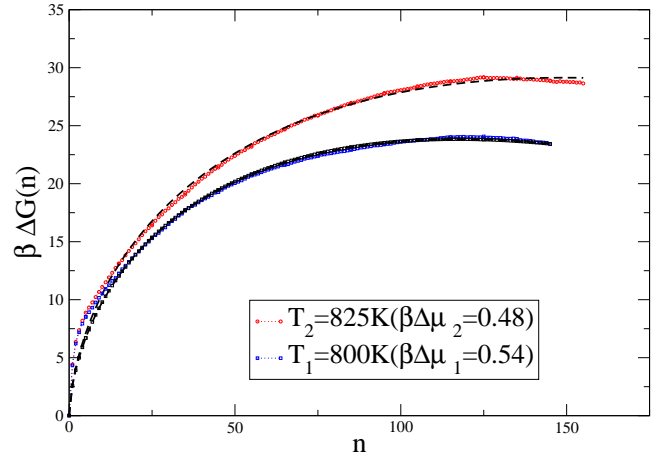}
\caption{Free energy barriers $\Delta G$ as a function of
the nucleus size $n$ for $T_{1}$=800K$(\beta\Delta\mu_{1}=0.54)$
and $T_{2}$=825K$(\beta\Delta\mu_{2}=0.48)$. Error-bars on $\beta
\Delta G$ are of the order of 1 $k_{B}T$. The dashed curves are
fits to the functional form given by CNT.}
 \label{Barriers}
\end{center}
\end{figure}

We computed the free energy barrier for crystal nucleation at
$T_{1}$=800K and $T_{2}$=825K, corresponding to $\beta
\Delta\mu_{1}$=$0.54$ and $\beta \Delta\mu_{2}$=$0.48$. The values
of $\beta \Delta\mu$ were estimated numerically by thermodynamic
integration from the coexistence temperature
and free energies
reported by Anwar et
al. \cite{Jamshed}. Fig.\ref{Barriers} shows the computed
nucleation barriers as a function of $n$.

As expected, $\Delta G$ decreases with supersaturation. Around
T$<$750K the barrier gets sufficiently low that spontaneous
nucleation can take place on the time scale of a simulation.
 The size of the critical nucleus, $N_c$, was estimated according to a fit of the functional form of the CNT, and we found $N_c\approx 120$ ions at
$T_{1}$ and $N_c\approx 150$ ions at $T_{2}$.  Koishi et al.
\cite{Japanese} estimated $N_c=120-130$  ions at 640K and 740K,
which is surprising in view of the CNT prediction that the size of
the critical nucleus scales as $(\gamma/|\Delta\mu|)^3$.  If we
make the usual assumption~\cite{Kelton} that $\gamma\sim \Delta h$
and $\Delta\mu\approx (\Delta h/T_m)(T-T_m)$, where $\Delta h$ is
the entropy of fusion per ion pair, then we would expect that
$N_c\sim (1-T/T_m)^{-3} $ and we would predict that the critical
nucleus at 640K should contain a quarter as many particles as those at
800K.

Next, we consider the structure and shape of the critical nucleus.
\begin{figure}[h!]
\begin{center}
\includegraphics[width=7.0cm,height=7.0cm,clip=true]{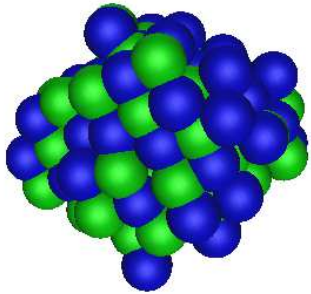}
\caption{Snapshot of the critical nucleus at $T_{2}$=825K:
the bulk NaCl structure is already evident. The critical crystal
nucleus seems to have rudimentary facets, which is in agreement
with the interpretation of the experiments in
ref.~\cite{NaClexp11}.} 
 \label{nucleus}
\end{center}
\end{figure}
Fig.\ref{nucleus} shows a snapshot of the critical nucleus at
$T_{2}$=825K. Note that the crystal presents rudimentary low-index
facets. In experiments \cite{NaClexp11,Balk} the existence of such
facets was postulated, as they may act as sites for subsequent
heterogeneous nucleation.

As can be seen from Fig.\ref{nucleus}, the critical nucleus
already shows the charge-ordered rock-salt structure of the bulk
phase. This indicates that, at least for NaCl, nucleation does not
proceed via an intermediate metastable phase~\cite{Ostwald}.
Fig.\ref{nucleus} also shows that, in the temperature range that
we studied, the critical nucleus is non-spherical.  In order to
quantify the degree of non-sphericity of the critical nucleus, we
expanded its density with respect to the center of mass in
rank-four-spherical-harmonics and constructed the quadratic
invariant $S_{4}$ \cite{WoldeS4}: we obtained $S_{4}(T_{1})$=0.115
and $S_{4}(T_{2})$=0.110. For a simple cube $S_{4}$=0.172 and for
a sphere $S_{4}$=0, therefore the shape of the critical nucleus is
closer to a cube than to a sphere. In other words, the critical
nucleus already exhibits the morphology of macroscopic NaCl
crystals~\cite{HerautMetois}.

In order to make sure that there were not finite size effects 
that resulted in 
interactions between image cluster, 
we visually checked that the critical nuclei didn't show a preferred spacial orientation and
that the minimum distance between them was bigger than half box. 

Moreover we calculated the Debye-Huckle screening lenght and found 
that it was smaller than 1 $\dot A$, showing that
the critical nuclei did not even electrostatically interact.
We could therefore conclude that there was no induced nucleation due to the
interaction between a critical cluster and its own periodic image.

Using the computed height of the nucleation barrier and the values
of  $\Delta\mu$ as input, we can estimate the surface free-energy
density $\gamma_{ls}$. To this end, we make use of the CNT
expression for the barrier height (eqn.~\ref{CNTBarrier}).
However, this expression assumes that the critical nucleus is
spherical. It is easy to derive the corresponding expression for a
cubical nucleus. The results for both estimates of $\gamma_{ls}$
are given in Table~\ref{tavola2}.  
\begin{table}[h!]
\centerline{
\begin{tabular}{||c|c|c||}
\hline {\bf T[K]} & $\gamma_{sphere}$ & $\gamma_{cube}$
\\ \hline {\bf 800} & $98\pm 2$ & $80\pm 1$
\\ \hline {\bf 825}   &  $99\pm 1$ & $79\pm 1$       \\ \hline {\bf
905}   &  $84.1^{(e)}$ & $67.8^{(e)}$\\ \hline
\end{tabular}}
\caption{ Surface free energy density (in $erg \ cm^{-2}$) assuming
spherical and cubical shape for critical nuclei. At 905K we report
the experimental value \cite{NaClexp11}. The entry in the lower
right-hand corner is based on the experimental estimate, but
assuming a cubical nucleus.}
\label{tavola2}
\end{table}
There exist experimental
estimates of $\gamma_{ls}$ at 905K~\cite{NaClexp11}. These
estimates are based on a somewhat questionable CNT expression for
the nucleation rate. Moreover, in ref.~\cite{NaClexp11} it is
assumed that the critical nucleus is spherical. The experimental
estimate of $\gamma_{ls}$ ($\gamma_{exp}$=$84.1[erg \ cm^{-2}]$) is
therefore not based on a direct determination. Nevertheless, in
the absence of other experimental data, this is the only number we
can compare to.

As the table shows, there is a fair agreement between simulation
and experiment. The experimental estimate for $\gamma$ is based on
the assumption that the critical nucleus is spherical. If it is
cubic, one would obtain the number in the lower right-hand corner.
In view of the many uncertainties in the analysis of the
experimental data, it is impossible to tell whether the
discrepancy between simulation and experiment is significant. An
estimate of $\gamma$ based on the experimental enthalpy of fusion
following Turnbull, would yield $\gamma\approx 115 [erg \ cm^{-2}]$
(see e.g.~\cite{Bartell}).

Huang et al. \cite{Bartell} estimated the solid-liquid
surface free-energy density of  NaCl from the nucleation rate
at 550K. To achieve this, Huang et al.  assumed that the CNT
expression for the nucleation rate is valid. Under those
assumptions, they obtained:  $\gamma$=$119.6 [erg \ cm^{-2}]$ for
a spherical nucleus.

In the present work, we can compute absolute nucleation rates
without making use of CNT. The only assumption we make is that the
barrier crossing is diffusive and that the
Zeldovitch (pre)factor is well approximated by the form given in
eqn.(\ref{eqn:rate}).  The Zeldovitch factors were found to be
respectively $Z_{1}$=0.016 for $T_{1}$ and $Z_{2}$=0.013 for
$T_{2}$. The true Zeldovitch factor may be slightly different, but
is in any event expected to be of $\mathcal{O}(10^{-2})$.  From
our MD simulations we obtained the following estimates for the
forward rates (eq. \ref{eqn:diff}) : $f^{+}_{crit}$=0.013
ps$^{-1}$ for $T_{1}$ and $f^{+}_{crit}$=0.033 ps$^{-1}$ for
$T_{2}$. Combining this information, we can compute the kinetic
prefactor of eq. \ref{eqn:prefactor}: $\kappa(T_{1})$= 6.9$\times
10^{36}$ $m^{-3}s^{-1}$ and $\kappa(T_{2})$= 1.5$\times 10^{37}$
$m^{-3}s^{-1}$. As is to be expected, the kinetic prefactor
depends only weakly on temperature. Using eqn.(\ref{eqn:rate}) we
then calculated the nucleation rate. The results are:
$I(T_{1})$=3$\times 10^{26\pm1}$ $m^{-3}s^{-1}$ and
$I(T_{2})$=4$\times 10^{24\pm1}$  $m^{-3}s^{-1}$.

These nucleation rates are about ten orders of magnitude higher
than the estimated experimental rate at 905K
($O(10^{16})$$m^{-3}s^{-1}$). Such a difference is hardly
surprising because the nucleation rate is expected to increase
rapidly with increasing supercooling.

We also computed the nucleation rate using an algorithm based on
the path-sampling techniques of refs.~\cite{FFS,TIS} (Forward-Flux
Sampling (FFS) and Transition-Interface Sampling (TIS)) . The
value obtained at $T_{1}=800$K is $I_{FFS-TIS}(T_{1}) =
O(10^{27\pm2})$ $m^{-3}s^{-1}$, which agrees surprisingly well
with the one obtained using the diffusive barrier-crossing
approach.

As the path-sampling method does not depend on the choice of the
reaction coordinate and does not require prior knowledge of the
phase space density, we can conclude that the method based on the
free-energy calculation gives us a good estimate for the
nucleation rate at this temperature.

We have also computed the nucleation rate at  T=750 K, using the
FFS-TIS method.  The computed nucleation rate is
$I_{FFS-TIS}(T=740K)$=$O(10^{35}$) $m^{-3}s^{-1}$. This is of the
same order of magnitude as the nucleation rate obtained calculated
 by  Koishi et al. \cite{Japanese} using ``brute-force'' MD. We
 cannot use the diffusive barrier crossing method at this
 temperature, as the barrier is too low to avoid spontaneous
 nucleation during long runs. However, if we assume that the kinetic
pre-factor and the surface free-energy density do not vary much
with temperature, we can use CNT to
 extrapolate the nucleation rate from $800 K$ to $740 K$. We find:
$I_{extrap}(T=740K) = O(10^{30}$) $m^{-3}s^{-1}$ which is
considerably lower than the results of the direct calculations.
This suggests that an extrapolation procedure based on CNT is not
reliable. A summary of our numerical results for the nucleation
barriers and rates are given in Table~\ref{tavola1}.
\begin{table}[h!]
\centerline{
\begin{tabular}{||c|c|c|c|c|c||}
\hline
 {\bf T[K]}& $\beta \Delta \mu $  &  $\beta \Delta G_{crit}$
& $f^{+}_{ncrit}$ & $I[m^{-3}s^{-1}]$ & $I_{FFS-TIS}[m^{-3}s^{-1}]$
\\ \hline {\bf 800}   & $0.54$ & $24$ &$0.013$
& 3 $\times 10^{26\pm1}$ & $10^{27\pm2}$
\\\hline {\bf 825}   &  $0.48$ & $29$ & $0.033$ &
4$\times 10^{24\pm1}$ & $--$   \\\hline
\end{tabular}}
\caption{Summary of the simulation results for the calculation of
the free-energy barrier and the nucleation rate for Tosi-Fumi
NaCl.}
 \label{tavola1}
\end{table}

A similar problem occurs if we try to extrapolate our numerical
data at $800 K$ and $825 K$ to $905 K$, the temperature of the
experiments of ref.~\cite{NaClexp11}. If  we can extrapolate our
simulation results to the experimental temperature of 905 K
($\beta \Delta\mu$=0.3) we obtain an estimated nucleation at 905 K
that is $O(6 \times 10^{11})$$m^{-3}s^{-1}$. This is some  nearly five
orders of magnitude less than the experimentally observed rate.

The discrepancy between simulation and experiment can be due to
several reasons. a) There might be an appreciable (but
unspecified) error in the experimental estimates (e.g. due to
residual heterogeneous nucleation). b) The estimated error in the
computed melting temperature of the Tosi-Fumi model is
$\pm$20K~\cite{Jamshed}. Such an uncertainty again easily
translates into a variation of the nucleation rate by several
orders  of magnitude. c) In view of the extreme sensitivity of
nucleation rates to the details of the intermolecular potential
(see, e.g.~\cite{AuerPoon}), the Tosi-Fumi potential may be
inadequate to model nucleation in NaCl, even though it can
reproduce the static properties of the solid and liquid NaCl
~\cite{Zykova}. d) Finally, it it is not quite correct to assume
that the kinetic prefactor, the surface free energy and the latent
heat of fusion are temperature-independent.

We can also compare our calculated nucleation rate at $800 K$ to
the rate estimated with CNT. In order to do that, we need to
compute the kinetic pre-factor $\kappa_{CNT}$ that, using the CNT
approximations, is \cite{Kelton}:
\begin{eqnarray}\label{eqn:prefactorCNT}
\kappa_{CNT} = Z \; \rho_{liq} \; \frac{24 D_{S} n_{crit}^{2/3} }{\lambda^{2}}.
\end{eqnarray}
The attachment rate of particles to the critical nucleus
($f^{+}_{n^{crit}}$) takes into account the number of available
attachment sites on the surface of a spherical nucleus
($n_{crit}^{2/3}$) and depends on the jump frequency for bulk
diffusion ($D_{S} /\ \lambda^{2}$), where $\lambda$ is the atomic
jump distance. Since the functional form of the nucleation barrier
can be fitted to the corresponding CNT expression, the computed
Zeldovitch factor (Z) coincides with the predicted one. We
computed the self-diffusion coefficient with MD simulations using
the $DL\_POLY$ package \cite{DLPOLY} in the supercooled liquid at
$T_{1}$=800K. We found $D^{Na}_{S} = 3.4 \times 10^{-5} cm^{2}
s^{-1}$, in good agreement with an estimate ($D^{Na}_{S} = 2.3
\times 10^{-5} cm^{2} s^{-1}$) based on extrapolation of the
available experimental data of Ref.~\cite{diffexp} to the
temperature $T_{1}$.  Since $D^{Na}_{S}/D^{Cl}_{S} \approx 1$, we
only considered the self-diffusion of the Na$^+$ ions. We
estimated $\lambda$ as a fitting parameter from the
$f^{+}_{n^{crit}}$ previously calculated; we obtained
$\lambda(T_{1})= 10^{2}\dot{A} $. However, considering that the
ion size is $\sigma_{Na} \sim 1.1$ $\dot{A}$,  this value for the
jump distance  seems unphysical  ($\lambda \sim 100 \sigma$).
Typically, one would expect $\lambda$ to be of the order of a mean
free path. In a molten salt, the mean free path of an ion is
certainly less than a particle diameter. This discrepancy also
suggests that the CNT picture is inadequate to describe crystal
nucleation of NaCl.

In summary, we have computed the crystal nucleation rate of sodium
chloride from the melt using two independent methods: one based on
calculations of the free-energy barrier and the diffusive
barrier-crossing and another based on a path-sampling approach. We
have found that, to within an order of magnitude, the two
approaches yield the same value for the nucleation rate. When we
use Classical Nucleation Theory to extrapolate our numerical data
to lower temperatures, we observe serious discrepancies with the
results of direct calculations. When we use CNT to extrapolate to
high temperatures, we find serious discrepancies with the
nucleation rates  found in experiments. Several factors may
contribute to this discrepancy but, at present, it is not yet
known which factor is most important.

\section*{Acknowledgments}

The work of the FOM Institute is part of the research program of
FOM and is made possible by financial support from the Netherlands
Organization for Scientific Research (NWO). C.V. gratefully
acknowledges the financial support provided through the European
Community Human Potential Program under contract
HPRN-CT-1999-00025, (Nucleus). E.S. gratefully acknowledges the
Spanish government for the award of a FPU Ph.D. grant, and the FOM
Institute for the hospitality during the period in which this work
was carried out.

C.V. and E.S. thanks Angelo Cacciuto and Rosalind Allen for
valuable discussions and suggestions, and Georgios Boulougouris
and Josep P\'amies for a critical reading of the manuscript.

\appendix
\renewcommand{\theequation}{A-\arabic{equation}} 
\setcounter{equation}{0} 
\section{Identification of crystalline clusters}
To distinguish between  solid-like and liquid-like particles and
 identify the particles belonging to a solid cluster,
we used the local bond order parameter introduced by Ten Wolde el al.~\cite{tenwolde}. 
Although the method we used is the same as the one proposed by Ten Wolde,
 the definition of a 'solid-like' particle is not rigorously the same. 
The q vector and the thresolds selected were optimazed for the NaCl Tosi-Fumi model.  

First we computed a normalized  complex vector $q_4$ for every particle $i$. 
Each component of this vector was given by:
\begin{eqnarray}\label{eqn:q4q4}
\vec{q}_{4,m}(i) = \frac{  \frac{1}{N_{b}(i)} \sum^{N_{b}(i)}_{j} \Upsilon_{4,m} 
(\theta_{i,j},\phi_{i,j})}{\vec{q}_{4,m}(i) \cdot  \vec{q}^{\ \ast}_{4,m}(i)} ,  \hspace{1 cm} m=[-4,4]
\end{eqnarray}
Where $N_b(i)$ is the number of neighbours of the particle $i$ within a cut-off radious of 4$\dot{A}$ 
(the first minimum in the Na-Cl radial distribution function). 

Then we computed a scalar product $q_4(i) \cdot q_4^{*}(j)$ for every particle $i$ 
with each of its neighbours particle $j$. 
A particle was considered to be 'solid-like' when at least 6 
of the scalar products were bigger than $0.35$.
Finally two 'solid-like' particles were considered 
to be neighbours in the same cluster if they were closer than 3.4 $\dot{A}$.

\begin{figure}[h!]
\begin{center}
\includegraphics[height=8cm,angle=-90,clip=true]{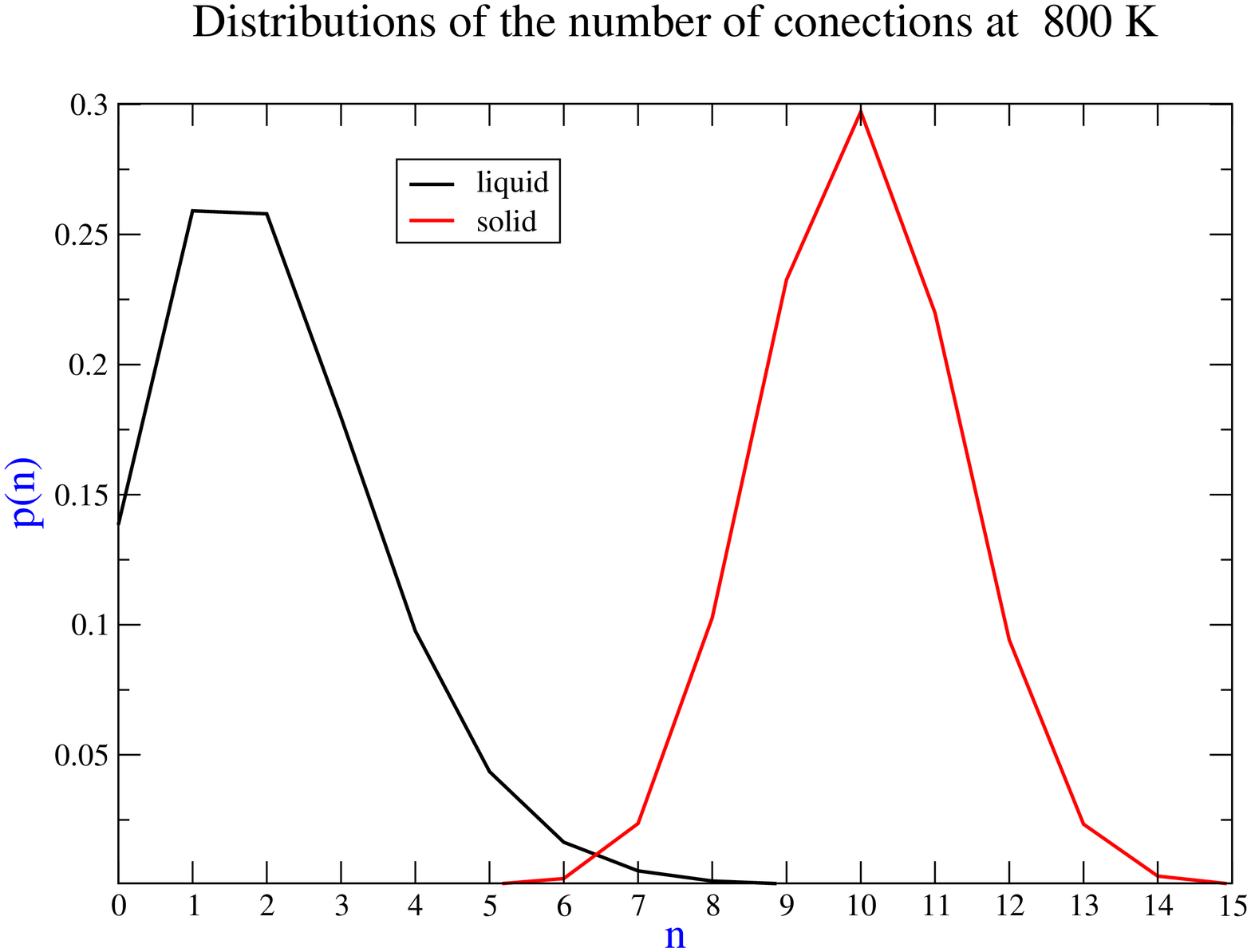}
\caption{Distributions of the number of scalar products bigger than $0.35$ at T=800 K
for the solid and for the liquid.}
\label{dist}
\end{center}
\end{figure}

Ionic fluids are more ordered than Lennard-Jones or Hard Spheres ones, as the radial
distribution function shows. 
However, with the method implemented by Ten Wolde, we were able to clearly distinguish between solid-like and liquid-like particles
in the NaCl Tosi-Fumi model.
We enclose a plot\ref{dist} that shows the distributions of the number of scalar products bigger than $0.35$ at T=800 K
for the solid and for the liquid: for values bigger than 6 a particle was considered to be solid-like.

\section*{References}

\bibliography{valeriani}

\end{document}